\documentclass[acmtog, nonacm]{acmart}

\let\oldmaketitle\maketitle
\renewcommand{\maketitle}{%
  \oldmaketitle%
  \thispagestyle{plain}%
  \pagestyle{plain}}

\setcopyright{none}                              %
\renewcommand\footnotetextcopyrightpermission[1]{}%
\usepackage{todonotes}
\usepackage{enumitem} 
\usepackage{algorithm}
\usepackage{algpseudocode}
\usepackage[T1]{fontenc}
\usepackage{xcolor}
\usetikzlibrary{arrows.meta,backgrounds,fit,positioning,shapes.geometric}
\usepackage{tikz}
\usepackage{amsmath,booktabs,tabularx}
\usepackage{xspace}
\usepackage[shortcuts]{extdash}
\usepackage{pgfplots}
\pgfplotsset{compat=1.18}
\usepackage[most]{tcolorbox}

\let\oldunderscore\_
\renewcommand{\_}{\oldunderscore\allowbreak}

\definecolor{ink}{HTML}{20252B}
\definecolor{muted}{HTML}{5B6570}
\definecolor{panel}{HTML}{F5F6F7}
\definecolor{component}{HTML}{E9EDF1}
\definecolor{watt}{HTML}{E7F1E3}
\definecolor{wattline}{HTML}{4A7C3C}
\definecolor{emap}{HTML}{E5EDF8}
\definecolor{emapline}{HTML}{3569A8}
\definecolor{inputfill}{HTML}{F4EEE4}
\definecolor{inputline}{HTML}{8A6E3B}

\newcommand{\escape}{ESCAPE\xspace}
\newcommand{\wn}{Wattnet\xspace}
\newcommand{\wnshort}{WN\xspace}
\newcommand{\emaps}{ElectricityMaps\xspace}
\newcommand{\emapshort}{EM\xspace}

\def\carbonunit{g$\cdot$CO$_2$\allowbreak eq/kWh\xspace}

\begin{document}
\nocite{wuCarbonEdgeLeveragingMesoscale2025b}
\title{Investigating the Impact of  CI-Source Sensitivity in Carbon-Aware Resource Management Systems}

\author{Gonçalo Ferreira}
\orcid{0000-0002-2819-1879}
\affiliation{%
  \institution{Informatics Institute, University of Amsterdam}
  \city{Amsterdam}
  \country{The Netherlands}
}
\email{g.j.teixeiradepinhoferreira@uva.nl}

\author{Jaime Iglesias Blanco}
\orcid{0009-0007-7761-5600}
\affiliation{%
  \institution{Instituto de Física de Cantabria (IFCA), CSIC-UC}
  \city{Santander}
  \postcode{39005}
  \country{Spain}
}
\email{iglesias@ifca.es}

\author{María Castrillo}
\orcid{0000-0001-7033-6907}
\affiliation{%
  \institution{Instituto de Física de Cantabria (IFCA), CSIC-UC}
  \city{Santander}
  \postcode{39005}
  \country{Spain}
}
\email{castrillo@ifca.unican.es}

\author{Álvaro López García}
\orcid{0000-0002-0013-4602}
\affiliation{%
  \institution{Instituto de Física de Cantabria (IFCA), CSIC-UC}
  \city{Santander}
  \postcode{39005}
  \country{Spain}
}
\email{aloga@ifca.es}

\author{Li Wu}
\orcid{0000-0001-7115-1517}
\affiliation{%
  \institution{Department of Computing Science, University of Alberta}
  \city{Edmonton, Alberta}
  \country{Canada}
}
\email{lwu15@ualberta.ca}

\author{Shashikant Ilager}
\orcid{0000-0003-1178-6582}
\affiliation{%
  \institution{Informatics Institute, University of Amsterdam}
  \city{Amsterdam}
  \country{The Netherlands}
}
\email{s.s.ilager@uva.nl}



\begin{abstract}
The sustainable growth of data centres increasingly depends on  Resource-Management Systems (RMSs) that aim to reduce carbon emissions of computing workloads without degrading performance. This need is intensified by the rise of Artificial Intelligence (AI)  workloads, which demand  exponentially growing amounts of compute and energy across geographically distributed infrastructure. Recent carbon-aware RMSs  exploit temporal and spatial variation in grid Carbon Intensity (CI) to shift workloads towards lower-carbon periods or locations.
These systems assume that the CI signal they consume is accurate enough for decision-making. In practice, however, CI estimates and forecasts can be inaccurate and uncertain, and CI providers' methods differ substantially. Such discrepancies may have a significant impact on RMS placement decision-making, raising concerns about the \textit{fragility} of existing solutions. Thus, in this work, we ask how discrepancies between plausible CI estimates propagate into resource-management decisions and operational outcomes. To answer this question, we harmonise and characterise CI traces from \textit{\wn} and \textit{\emaps} across five Central European zones. We then propose our evaluation framework: External CI Signal Characterisa\-tion And Perturbation Evaluation  (\textit{\escape}), which assesses provider switching and controlled CI perturbations in a reproduced placement setting, including empirical perturbations derived from observed disagreement between CI-sources. We implement our framework using CarbonEdge as a use case, and measure changes in decision outputs, such as site rankings and placements, and operational outcomes, such as carbon proxy, energy use, and latency. Our results show that switching CI-source  changes the dominant placement site in \(43.50\%\) cases, and alters reported operational outcomes. These changes occur despite high rank correlation and structured disagreement between CI-sources. This observed sensitivity to CI-source choice motivates robustness-aware carbon-aware placement, in which CI uncertainty should be explicitly accounted for in resource-allocation decisions.
\end{abstract}

\maketitle

\section*{Availability of Data and Material}
\noindent
The source code and data supporting this study are available at
\url{https://github.com/g-uva/escape-ci}.

\section{Introduction}

The environmental footprint of data centers has become a first-order concern in recent years. Large-scale computing infrastructures, including hyperscale cloud platforms, High-Performance Computing (HPC) facilities, and distributed edge systems, support critical scientific, commercial, and user-facing workloads~\cite{hanafyCarbonScalerLeveragingCloud2023,souzaCASPERCarbonAwareScheduling2024,wuCarbonEdgeLeveragingMesoscale2025b}.
Recent progress in Artificial Intelligence (AI) has further intensified this pressure, where training and inference require large amounts of computational resources and power ~\cite{pattersonCarbonEmissionsLarge2021a,luccioniPowerHungryProcessing2024}. Estimates indicate that data centers consumed approximately $415$ terawatt-hours ($TWh$) of electricity in 2024, which corresponds to 1.5\% of global electricity consumption~\cite{ieaEnergyAI2025}, with a significant Carbon Footprint (CFP).

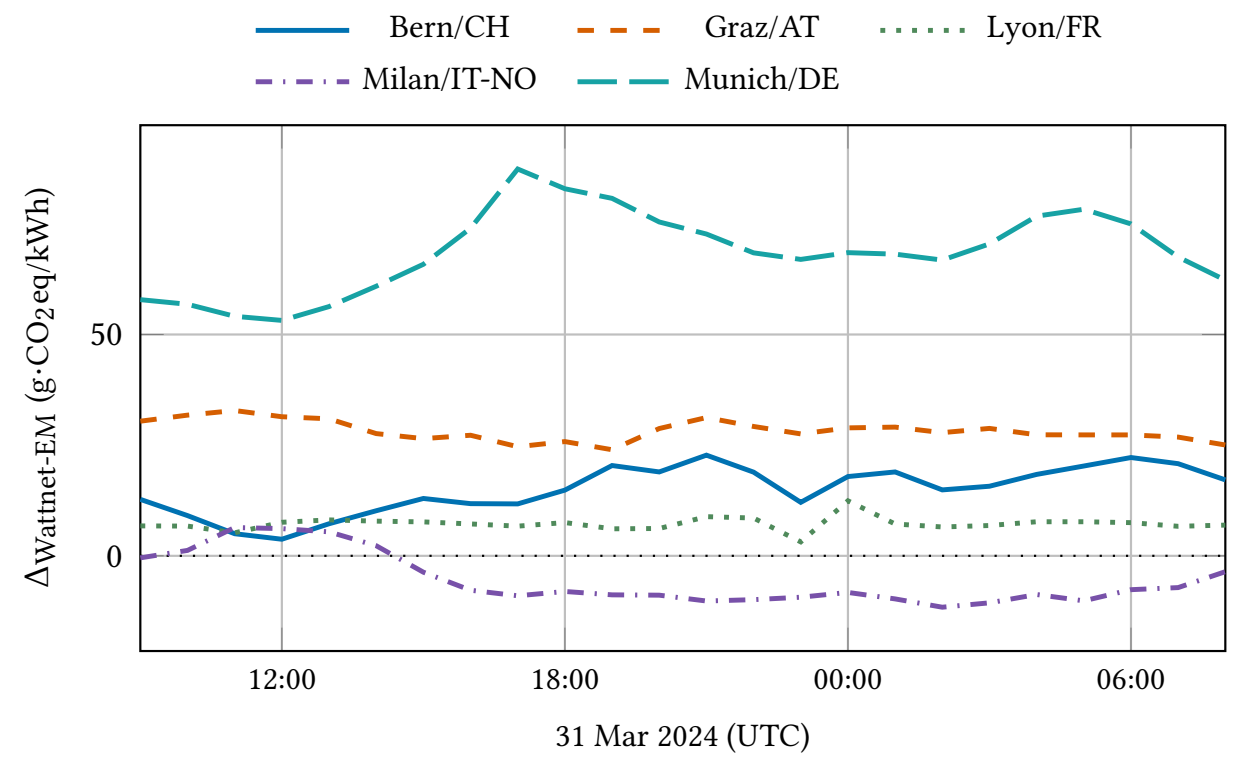
\begin{figure}[t]
    \centering
    \resizebox{\columnwidth}{!}{
\definecolor{cepalette0}{HTML}{0072B2} 
\definecolor{cepalette1}{HTML}{D55E00} 
\definecolor{cepalette2}{HTML}{4F8A5B} 
\definecolor{cepalette3}{HTML}{7851A9} 
\definecolor{cepalette4}{HTML}{17A2A4} 
\begin{tikzpicture}
\begin{axis}[
  width=\columnwidth,
  height=0.58\columnwidth,
  xmin=0,
  xmax=23,
  xtick={3,9,15,21},
  xticklabels={12:00,18:00,00:00,06:00},
  xticklabel style={font=\scriptsize},
  yticklabel style={font=\scriptsize},
  label style={font=\scriptsize},
  legend columns=3,
  legend style={
      font=\scriptsize,
      draw=none,
      at={(0.5,1)},
      anchor=south,
      yshift=3pt,
      inner sep=0pt,
      /tikz/every even column/.append style={column sep=0.55em}
  },
  ylabel={$\Delta_{\mathrm{\text{\wn-\emapshort}}}$ (\carbonunit)},
  xlabel={31 Mar 2024 (UTC)},
  grid=major,
]
\addplot[black, dotted, line width=0.4pt, forget plot] coordinates {(0,0) (23,0)};
\addplot[cepalette0, solid, line width=0.9pt] coordinates {(0,12.7554) (1,9.08908) (2,4.99482) (3,3.75006) (4,7.24456) (5,10.1831) (6,12.9577) (7,11.7886) (8,11.7313) (9,14.8448) (10,20.398) (11,18.9516) (12,22.7425) (13,18.905) (14,12.084) (15,17.9095) (16,18.9503) (17,14.8953) (18,15.721) (19,18.3922) (20,20.2886) (21,22.2068) (22,20.8167) (23,17.1535)};
\addlegendentry{Bern/CH}
\addplot[cepalette1, dashed, line width=0.9pt] coordinates {(0,30.416) (1,31.7829) (2,32.8243) (3,31.4224) (4,30.9525) (5,27.6145) (6,26.4793) (7,27.2398) (8,24.6498) (9,25.8101) (10,23.9666) (11,28.7584) (12,31.2378) (13,29.2257) (14,27.5448) (15,28.9079) (16,29.0787) (17,27.8298) (18,28.7865) (19,27.3444) (20,27.3199) (21,27.3156) (22,26.8225) (23,25.0389)};
\addlegendentry{Graz/AT}
\addplot[cepalette2, dotted, line width=0.9pt] coordinates {(0,6.76165) (1,6.7187) (2,5.19988) (3,7.57872) (4,8.11771) (5,7.83529) (6,7.66557) (7,7.20507) (8,6.70482) (9,7.51476) (10,6.10579) (11,6.17322) (12,8.85042) (13,8.53817) (14,3.14876) (15,12.4316) (16,7.18783) (17,6.52305) (18,6.84617) (19,7.70073) (20,7.69966) (21,7.4969) (22,6.67181) (23,6.94108)};
\addlegendentry{Lyon/FR}
\addplot[cepalette3, dashdotted, line width=0.9pt] coordinates {(0,-0.454198) (1,1.22481) (2,6.39515) (3,6.14305) (4,5.41736) (5,2.28468) (6,-3.69685) (7,-7.77321) (8,-9.02292) (9,-8.05082) (10,-8.81476) (11,-8.88802) (12,-10.2062) (13,-9.89868) (14,-9.34824) (15,-8.269) (16,-9.73033) (17,-11.6055) (18,-10.6166) (19,-8.73888) (20,-10.1667) (21,-7.64734) (22,-7.17381) (23,-3.58008)};
\addlegendentry{Milan/IT-NO}
\addplot[cepalette4, dash pattern=on 7.2pt off 2.25pt, line width=0.9pt] coordinates {(0,57.8723) (1,56.8286) (2,54.1093) (3,53.184) (4,56.3063) (5,60.9022) (6,65.8624) (7,74.0098) (8,87.3812) (9,82.9234) (10,80.7541) (11,75.4206) (12,72.6947) (13,68.4433) (14,66.9638) (15,68.4907) (16,68.1523) (17,66.8121) (18,70.4671) (19,76.7336) (20,78.2946) (21,74.9779) (22,67.4987) (23,62.2773)};
\addlegendentry{Munich/DE}
\end{axis}
\end{tikzpicture}}
    \caption{Hourly difference between the Carbon Intensity (CI) estimates reported by \wn and \emapshort for five Central European locations encompassing 24 hours between 31 March and 1 April 2024. The reference line (\(0\)) from EM serves as the reference for the \wn $\Delta$. The absolute difference reaches \(87.4\)~\carbonunit at its peak (Munich/DE).}
    \label{fig:intro-provider-disagreement}
\end{figure}

A growing body of work addresses this data center sustainability problem through carbon-aware Resource-Management Systems (RMSs), deciding \textit{when} and \textit{where} computational workloads should execute. To inform these decisions, these RMSs estimate carbon emissions by combining energy consumption with environmental signals, such as the grid Carbon Intensity (CI) signal~\cite{wiesnerLetsWaitAwhile2021,radovanovicCarbonAwareComputingDatacenters2023,hanafyCarbonScalerLeveragingCloud2023,souzaCASPERCarbonAwareScheduling2024,wuCarbonEdgeLeveragingMesoscale2025b}.
The CI quantifies the greenhouse gas emissions associated with each unit of electricity, enabling lower-carbon operational decisions.
Because CI varies across both \textit{time} and \textit{location}, systems that use CI mainly exploit \textit{temporal} and \textit{spatial} flexibility of workloads' computation to decrease the CFP~\cite{asadovCarbonAwareSpatioTemporalWorkload2025}.

However, a major limitation shared by current carbon-aware RMS approaches is their dependence on an external CI signal provider, where the signal data is not observed but estimated using provider-specific data sources, emission factors, spatial boundaries, and flow-tracing methods. Moreover, future CI values depend on forecasting models; both estimation and prediction models produce uncertainty which can propagate into downstream decision differences~\cite{yanEnsembleCIEnsembleLearning2025,liUncertaintyAwareDecarbonizationDatacenters2025}. Most studies in literature obtain CI values from sources like \emapshort ~\cite{sukprasertLimitationsCarbonAwareTemporal2024,wuCarbonEdgeLeveragingMesoscale2025b,radovanovicCarbonAwareComputingDatacenters2023}, and WattTime ~\cite{dodgeMeasuringCarbonIntensity2022a,buchananCarbonAwareComputing2023}. \emapshort is widely used because of its broad geographical coverage and accessible CI data. Building on these capabilities, \wn (\wnshort) provides a more recent Europe-focused alternative, using cross-border electricity-flow tracing to produce high-resolution carbon- and water-intensity estimates.

A key concern is, however, the substantial discrepancies between CI sources. As illustrated in Figure~\ref{fig:intro-provider-disagreement}, the hourly difference across the analysed Central European locations reaches \(87.4\)\,\carbonunit (31 March 2024); equivalent to $33.1\%$ of the corresponding \emapshort base-value, which may be non-negligible for CI-sensitive RMS decisions. Moreover, existing work lacks an accessible, end-to-end method for measuring CI-source discrepancies into placement and operational outcome differences, and translating it into empirically-grounded stress tests across RMSs in order to measure operational outcomes such as performance, latency, and carbon. This gap leaves their \textit{fragility} insufficiently characterised, since plausible CI variations may occur and alter placement decisions and operational outcomes.

{\looseness=-1
Accordingly, in this paper, we ask \textit{how sensitive are placement decisions and operational outcomes to differences in CI estimation and prediction?} We study this by harmonising and characterising the differences between the \emapshort and \wn CI traces, measuring downstream placement propagation of source-switching, and analysing decision robustness by introducing synthetic and empirically-grounded perturbations. We evaluate \textit{decision-level sensitivity} through changes in site rankings and placements, and \textit{operational outcomes} through changes in the carbon proxy,~energy use,~and~\mbox{latency}.\par}

We implement our framework instantiating CarbonEdge~\cite{wuCarbonEdgeLeveragingMesoscale2025b} as use case, an RMS from literature that develops carbon-aware workload scheduling for mesoscale edge systems. Our results show that although CI rankings remain strongly correlated (0.970), switching from \emapshort to \wn changes the top-ranked CI site in \(14.60\%\) of 2,486 aligned decision steps per run pair over 226 days. The dominant placement site (the site receiving the most application assignments) changes in \(43.50\%\) of the cases. Mean energy use and latency decrease by  \(1.02\%\) and \(0.48\%\), respectively, while the operational carbon proxy increases by \(15.07\%\). These results show that CI-source sensitivity should be evaluated through its decision-level consequences, as large discrepancies may preserve decisions whereas small differences may reverse candidate rankings. Carbon-aware RMSs should account for this CI-source disagreement during its design and evaluation.

The remainder of this manuscript includes Sections~\ref{sec:background} and~\ref{sec:related-work} which introduce CI estimation, carbon-aware resource management, and related work. Section~\ref{sec:framework} presents the framework, while Section~\ref{sec:experimental-evaluation} describes the experimental setup and metrics. Section~\ref{sec:results-analysis} analyses CI-source agreement, decision sensitivity, and operational outcomes. Finally, Section~\ref{sec:conclusion} summarises the findings, limitations, and recommendations.

\section{Background and Motivation}
\label{sec:background}

\subsection{Wattnet}

Wattnet~\cite{CASTRILLOMELGUIZO2026101473} is an open-source tool and public service that estimates the carbon and water footprint of electricity consumption across Europe at 15-minute resolution, with a 72-hour forecast of both footprints. It ingests generation by production type and cross-border physical flows from the ENTSO-E Transparency Platform~\cite{entsoeTransparencyPlatform}, complemented by NESO~\cite{nesoDataPortal}/Elexon~\cite{elexonInsightsSolution} for Great Britain and EPİAŞ~\cite{epiasTransparencyPlatform} for Türkiye, and covers 60 zones, of which 51 currently have data. Each node of the network is the finest geographical unit available (country, bidding zone, or control area).

Before any footprint is computed, a quality-control stage detects outliers in the operator data and backfills missing values with estimates. The electricity consumed in each zone is then attributed to its generators using flow tracing with proportional sharing over the grid topology, including successive and looped exchanges. The resulting mix is combined with operational or life-cycle emission factors taken from institutional sources and peer-reviewed literature (IPCC, UNECE, among others) to obtain the carbon intensity of the consumed electricity (the \textit{global} footprint), as opposed to the \textit{local} footprint, which only accounts for domestic generation. Hydro pumped storage is excluded from the local calculation, since the footprint of its output depends on the unobserved charging history of the reservoir.

Wattnet is distributed both as open-source software that can be deployed on premises and as a public service accessible through a REST API, a dashboard, and a Python client. This makes the methodology transparent and reproducible, which is the property that allows us to characterise the sensitivity of resource-management decisions to the CI source.

\subsection{CI Estimation Approaches (Wattnet vs. \emapshort)}

Both providers estimate location-based, consumption-oriented CI from the same family of inputs: generation per production type and physical cross-border flows published by European transmission system operators, traced across borders with a flow-tracing algorithm, and weighted with per-technology emission factors~\cite{tranbergRealtimeCarbonAccounting2019,CASTRILLOMELGUIZO2026101473}. Both can be written in the same form. Let $G_{k,\mathrm{tech},t}$ be the generation of technology $\mathrm{tech}$ in zone $k$ at time $t$, and $c_{i\leftarrow k,t}$ the share of the electricity consumed in zone $i$ that was generated in zone $k$, obtained by flow tracing. The consumption-based CI of zone $i$ is
\begin{equation}
\mathrm{CI}_{i,t} = \sum_{\mathrm{tech}} s_{i,\mathrm{tech},t}\,\mathrm{EF}_{\mathrm{tech}},
\qquad
s_{i,\mathrm{tech},t} = \sum_{k} c_{i\leftarrow k,t}\,\frac{G_{k,\mathrm{tech},t}}{\sum_{\mathrm{tech}'} G_{k,\mathrm{tech}',t}},
\label{eq:ci}
\end{equation}
where $s_{i,\mathrm{tech},t}$ is the share of technology $\mathrm{tech}$ in the mix attributed to zone $i$ and $\mathrm{EF}_{\mathrm{tech}}$ its emission factor. Summing over technologies first gives the equivalent formulation in terms of zonal production intensities, $\mathrm{CI}_{i,t}=\sum_k c_{i\leftarrow k,t}\,\mathrm{CI}^{\mathrm{prod}}_{k,t}$.

The two providers therefore share the structure of Eq.~\eqref{eq:ci}, and the differences that remain lie in its inputs and implementation: (i) the emission factors $\mathrm{EF}_{\mathrm{tech}}$ and their scope (operational or life-cycle); (ii) the temporal resolution and the zoning, which determine the nodes of the traced network; (iii) the treatment of storage, incomplete data and outliers before tracing; and (iv) coverage and replicability. Wattnet publishes its factors, algorithm, and quality-control criteria, whereas the \emapshort model is not fully reproducible from public information~\cite{CASTRILLOMELGUIZO2026101473}.

\section{Related Work}
\label{sec:related-work}
In this section, we provide an overview of the electricity CI signals used in carbon-aware computing, RMS methods used to act on them, and approaches that account for uncertainty in CI estimates and predictions. Together, these motivate the evaluation of how CI-source error and disagreement propagates into different placement decisions and operational outcomes.

\noindent \textbf{CI Sources:} CI signals combine electricity-system data with emis\-sions-accounting assumptions. Typically those values are not observed but estimated or predicted, using distinct calculation methods. The ENTSO-E Transparency Platform~\cite{entsoeTransparencyPlatform} is used as an upstream data source by CI providers.
One of such providers is ElectricityMaps (\emapshort), which calculates consumption-based CI using generation mixes, cross-border flows, and emission factors, with estimation procedures for missing observations and forecasting modules for future conditions~\cite{electricityMapsMethodology}.
Another provider is \wn (\wnshort) that similarly uses flow tracing to derive consumption-based footprints, while exposing operational and life-cycle accounting options through an open-source implementation~\cite{CASTRILLOMELGUIZO2026101473}.
Both \emapshort and \wn are used in this study's experiment.
WattTime, another widely used CI provider, provides marginal operating emissions signals, intended to estimate the emissions response to changes in electricity demand\footnote{\url{https://watttime.org/data-science/methodology-validation/}}. 
Those CI source providers address a different accounting question from average consumption-based CI to produce their signals. This helps explain the differences between them, but does not mitigate or quantify its uncertainty impact in RMS approaches that depend on it.

\noindent \textbf{Carbon-aware resource allocation:} Carbon-aware Resource Management Systems (RMSs) exploit \textit{temporal} or \textit{spatial} flexibility to reduce emissions, subject to constraints such as workload, infrastructure, Service-Level Objectives (SLOs), and external signals like~CI.
\textit{Cloud:} Cloud systems can combine flexible execution times, elastic source provisioning, and geographically distributed capacity. Prominent examples using CI include the Google's Carbon-Intelligent Compute Management study, which uses day-ahead forecast and risk-aware optimisation to impose carbon-aware capacity limits on flexible workloads~\cite{radovanovicCarbonAwareComputingDatacenters2023}; CarbonScaler, which exploits workload elasticity to adjust resource allocation under completion deadlines~\cite{hanafyCarbonScalerLeveragingCloud2023}; and CASPER, which jointly considers request routing and provisioning for distributed web services under latency objectives~\cite{souzaCASPERCarbonAwareScheduling2024}.
\textit{High-Performance Computing (HPC):} in HPC workloads, carbon-aware scheduling must account for parallel resource requirements and the performance consequences of delaying or relocating jobs.
Carbon-aware scheduling has also been explored in HPC that incorporate time-varying CI and predicted job energy requirements into HPC scheduling policies~\cite{benhariAdaptiveCarbonAwareScheduling2026}.
\textit{Edge:} Edge systems face tighter latency, communication, and device-resource constraints, which restrict the opportunities for workload shifting using CI.
Exploring these constraints, CarbonEdge exploits mesoscale spatial CI variation to place workloads across edge data centers while respecting latency and resource constraints~\cite{wuCarbonEdgeLeveragingMesoscale2025b}.
Previous research reviews workload shifting and propose a joint spatio-temporal approach for edge--cloud environments that uses CI signals for scheduling decisions while also accounting for embodied emissions~\cite{asadovCarbonAwareSpatioTemporalWorkload2025}.

Many carbon-ware studies directly rely on externally supplied CI estimates or forecasts for their scheduling or placement decision ~\cite{radovanovicCarbonAwareComputingDatacenters2023,hanafyCarbonScalerLeveragingCloud2023,souzaCASPERCarbonAwareScheduling2024,wuCarbonEdgeLeveragingMesoscale2025b}.
Reliance on uncertain CI estimates can constrain temporal and spatial workload shifting, even under favourable conditions~\cite{sukprasertLimitationsCarbonAwareTemporal2024}. Prior work has quantified the CI uncertainty and incorporated calibrated prediction bounds into scheduling to avoid emissions increases~\cite{liUncertaintyAwareDecarbonizationDatacenters2025}.
Other studies inject controlled errors into CI inputs, including randomly sampled deviations, to evaluate how prediction inaccuracies affect scheduling performance and carbon savings~\cite{bovornkeeratirojGreenThriftOptimizingCarbon2025, bostandoostDatadrivenAlgorithmSelection2025}.
These approaches motivate evaluating CI-source sensitivity through controlled perturbation experiments, combining synthetic changes that isolate specific effects with empirically derived perturbations that reflect the disagreement between CI source providers. In this study, we instantiate a CarbonEdge simulator into our framework as a controlled experimental setting to evaluate its ability to trace how CI uncertainty propagates into placement decisions and operational outcomes.

\section{Study Design and Methodology}
\label{sec:framework}

We introduce the \textbf{E}xternal CI \textbf{S}ignal \textbf{C}haracterisa\-tion \textbf{A}nd \textbf{P}ertur\-bation \textbf{E}valuation (\textit{\escape}) framework, a reusable and policy-agnostic methodology for systematically quantifying how uncertainty or misalignment in external CI signals can propagate into RMS decisions and downstream outcomes, including carbon, latency, and energy. The framework is policy-agnostic in that it separates signal harmonisation, signal characterisation, perturbation, and decision-level evaluation from the specific RMS policy under study. As a result, it can be applied to different decision mechanisms such as placement, routing, temporal shifting, scaling, or admission-control policies; provided that their decisions and outcomes can be replayed or evaluated under alternative CI inputs.

\subsection{System Model}

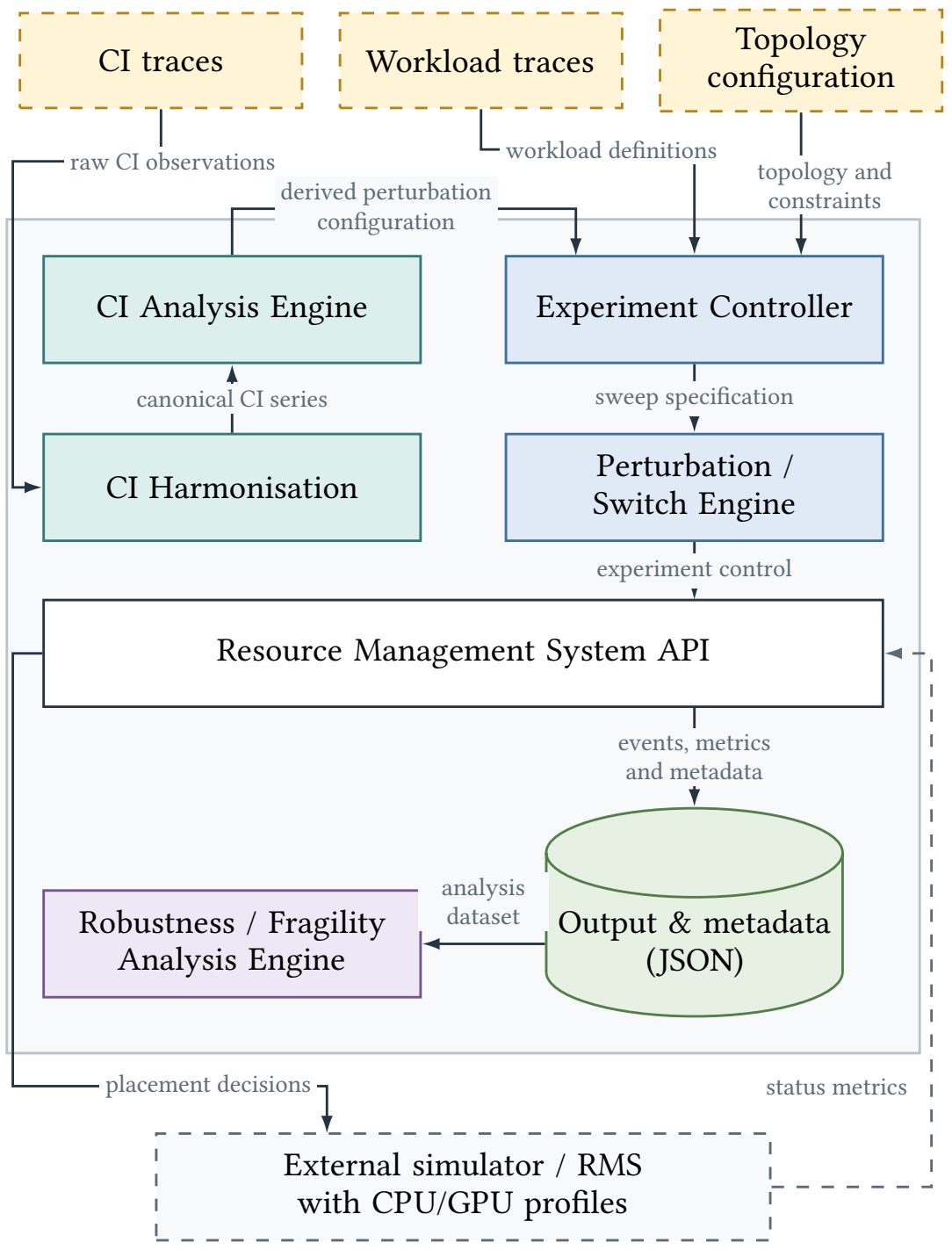
\begin{figure}[t]
  \centering
  \hspace*{-1,2cm}%
  \resizebox{\columnwidth}{!}{%
    \usetikzlibrary{arrows.meta,backgrounds,calc,fit,positioning,shapes.geometric}

\definecolor{ink}{HTML}{263442}
\definecolor{muted}{HTML}{5C6B77}
\definecolor{rule}{HTML}{B8C2CA}
\definecolor{panel}{HTML}{F7F9FA}
\definecolor{inputfill}{HTML}{FFF4D6}
\definecolor{inputline}{HTML}{B98218}
\definecolor{datafill}{HTML}{DCEFED}
\definecolor{dataline}{HTML}{347B78}
\definecolor{controlfill}{HTML}{DFEAF6}
\definecolor{controlline}{HTML}{3D6F9F}
\definecolor{analysisfill}{HTML}{EEE7F5}
\definecolor{analysisline}{HTML}{76558F}
\definecolor{storefill}{HTML}{E6F0E3}
\definecolor{storeline}{HTML}{557D4A}

\begin{tikzpicture}[
  font=\fontsize{8}{9}\selectfont,
  component/.style={rectangle, minimum height=9mm, text width=29mm,
    align=center, line width=.6pt, draw=ink, fill=white, inner sep=1.4mm},
  data module/.style={component,draw=dataline,fill=datafill},
  control module/.style={component,draw=controlline,fill=controlfill},
  analysis module/.style={component,draw=analysisline,fill=analysisfill},
  user input/.style={component,text width=21mm,minimum height=8mm,
    draw=inputline,fill=inputfill,dashed},
  external/.style={component,text width=49mm,draw=muted,fill=panel,dashed},
  store/.style={cylinder,shape border rotate=90,aspect=.30,
    minimum height=14mm,text width=23mm,align=center,inner sep=1mm,
    draw=storeline,fill=storefill,line width=.6pt},
  flow/.style={-{Latex[length=1.5mm,width=1mm]},draw=ink,line width=.6pt},
  return flow/.style={flow,draw=muted,dashed},
  label/.style={font=\fontsize{6}{7}\selectfont,text=muted,
    fill=panel,inner sep=.5mm,align=center},
  legend/.style={font=\fontsize{6}{7}\selectfont,text=ink,anchor=west}
]
\node[user input] (ci) at (0,0) {CI traces};
\node[user input] (workload) at (2.7,0) {Workload traces};
\node[user input] (topology) at (5.4,0) {Topology\\configuration};
\node[data module] (cianalysis) at (.6,-2.1) {CI Analysis Engine};
\node[data module] (harmonisation) at (.6,-3.6) {CI Harmonisation};
\node[control module] (orchestration) at (4.5,-2.1) {Experiment Controller};
\node[control module] (perturbation) at (4.5,-3.6) {Perturbation /\\Switch Engine};
\node[component,text width=68mm] (scheduler) at (2.55,-5.0) {Resource Management System API};
\node[analysis module] (robustness) at (.6,-7.45)
  {Robustness / Fragility\\Analysis Engine};
\node[store] (output) at (4.5,-7.45) {Output \& metadata\\(JSON)};
\begin{scope}[on background layer]
  \node[fit=(cianalysis)(harmonisation)(orchestration)(perturbation)
    (scheduler)(robustness)(output),inner xsep=3mm,inner ysep=3mm,
    draw=rule,fill=panel,line width=.6pt] (system) {};
\end{scope}
\node[external] (external) at (2.55,-9.5)
  {External simulator / RMS\\with CPU/GPU profiles};
\draw[flow] (ci.south) -- (0,-.85) -- (-1.25,-.85) |- (harmonisation.west);
\node[label,fill=white] at (.1,-.85) {raw CI observations};
\draw[flow] (harmonisation.north) -- (cianalysis.south)
  node[label,midway] {canonical CI series};
\draw[flow] (cianalysis.north) -- (.6,-1.25) -- (3.5,-1.25) -- (3.5,-1.65);
\node[label] at (1.9,-1.25) {derived perturbation\\configuration};
\draw[flow] (workload.south) -- (2.7,-.75) -- (4.5,-.75) -- (orchestration.north);
\node[label,fill=white] at (3.8,-.75) {workload definitions};
\draw[flow] (topology.south) -- (5.4,-1.65);
\node[label,fill=white] at (5.6,-1.05) {topology and\\constraints};
\draw[flow] (orchestration.south) -- (perturbation.north)
  node[label,midway] {sweep specification};
\draw[flow] (perturbation.south) -- (4.5,-4.55)
  node[label,midway] {experiment control};
\draw[flow] (4.5,-5.45) -- (output.north)
  node[label,midway] {events, metrics\\and metadata};
\draw[flow] (output.west) -- (robustness.east)
  node[label,midway,above=1mm,text width=10mm] {analysis\\dataset};
\draw[flow] (scheduler.west) -- (-1.25,-5.0) -- (-1.25,-8.65)
  -- (1.4,-8.65) -- (1.4,-9.05);
\node[label,fill=white] at (.4,-8.65) {placement decisions};
\draw[return flow] (external.east) -- (6.5,-9.5) -- (6.5,-5.0) -- (scheduler.east);
\node[label,fill=white] at (5.7,-8.65) {status metrics};
\end{tikzpicture}
  }
  \caption{Proposed \escape framework architecture and information flows.}
  \Description{Proposed \escape framework architecture and information flows.}
  \label{fig:architecture-diagram}
\end{figure}

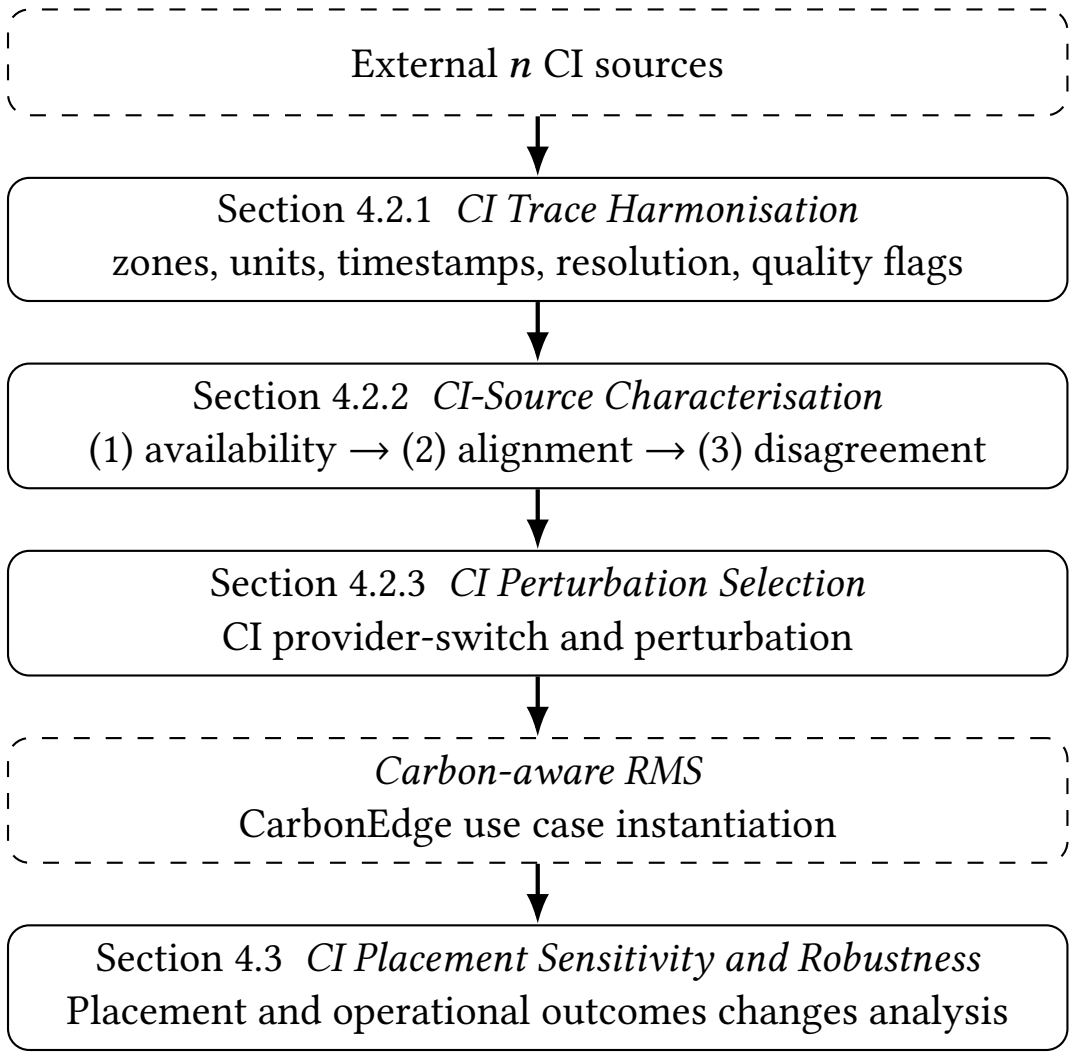
\begin{figure}[t]
    \centering
    \begin{tikzpicture}[
        node distance=4mm,
        box/.style={
            draw, rounded corners, align=center,
            minimum width=.82\columnwidth,
            minimum height=7mm, font=\small
        },
        arrow/.style={-{Latex[length=2mm]}, thick}
    ]
        \node[box, dashed] (sources) {External $n$ CI sources};
        \node[box, below=of sources] (harm) {
            Section~\ref{subsec:ci-harmonisation}\enspace
            \textit{\nameref{subsec:ci-harmonisation}}\\
            zones, units, timestamps, resolution, quality flags
        };
        \node[box, below=of harm] (char) {
            Section~\ref{subsec:ci-characterisation}\enspace
            \textit{\nameref{subsec:ci-characterisation}}\\
            (1) availability $\rightarrow$ (2) alignment
            $\rightarrow$ (3) disagreement
        };
        \node[box, below=of char] (select) {
            Section~\ref{subsec:ci-perturbation-selection}\enspace
            \textit{\nameref{subsec:ci-perturbation-selection}}\\
            CI provider-switch and perturbation
        };
        \node[box, dashed, below=of select] (system) {
            \textit{Carbon-aware RMS}\\
            CarbonEdge use case instantiation
        };
        \node[box, below=of system] (outcomes) {
            Section~\ref{subsec:decision-level-ci-analysis}\enspace
            \textit{\nameref{subsec:decision-level-ci-analysis}} \\
            Placement and operational outcomes changes analysis
        };

        \draw[arrow] (sources) -- (harm);
        \draw[arrow] (harm) -- (char);
        \draw[arrow] (char) -- (select);
        \draw[arrow] (select) -- (system);
        \draw[arrow] (system) -- (outcomes);
    \end{tikzpicture}
    \caption{Workflow connecting CI-source characterisation to
    provider-grounded perturbation and decision-level robustness analysis.}
    \label{fig:ci-workflow}
\end{figure}

Figure~\ref{fig:architecture-diagram} presents the architecture and information flows of the proposed \escape framework. Its architecture allows the framework to act as a policy-agnostic layer around an RMS under test, leaving its internal scheduling or placement logic unchanged. It accepts CI traces, workload traces, and a topology configuration as external inputs. Its components harmonise and compare the CI signals, configure the experiments, and analyse the resulting decision robustness. The RMS API component provides an interface for an external simulator or RMS, providing alternative CI inputs and collecting placement decisions and operational metrics for comparison. In this study, this is used by instantiating the CarbonEdge's use-case (Section \ref{sec:experimental-evaluation}). The following sub-sections describe the four workflow stages, referencing Figure~\ref{fig:ci-workflow}.

\subsection{CI Harmonisation, Characterisation, and Perturbation Selection}
\label{sec:ci-method}

\begin{table}[t]
\centering
\caption{Characterisation-to-perturbation selection rules.}
\label{tab:perturbation-selection}
\small
\begin{tabularx}{\columnwidth}{@{}X X@{}}
\toprule
\textbf{Condition} & \textbf{Effect} \\
\midrule
Sufficient co-available CI observation across providers and candidate sites
& Include provider-switch experiments. \\

Material site-specific CI offset
& Include additive perturbations. \\

Material proportional difference between providers' CI values
& Include multiplicative perturbations. \\

Material or correlated post-affine residuals
& Include empirical residual block resampling. \\

Material CI disagreement in both affine and residual components
& Include joint affine--residual perturbations. \\
\bottomrule
\end{tabularx}
\end{table}

\subsubsection{CI Trace Harmonisation}
\label{subsec:ci-harmonisation}
We align provider-specific zones, timestamps, units, and temporal resolution while retaining quality flags and missing observations. This enables comparable CI inputs without treating representational differences as uncertainty or assuming equivalent estimation methods.

\subsubsection{CI-Source Characterisation}
\label{subsec:ci-characterisation}
We assess three main properties: \emph{coverage}, which identifies shared spatial and temporal availability; \emph{numerical alignment}, which compares CI values on matched observations without assuming ground truth; and \emph{disagreement structure}, which separates site-specific offsets and scaling from remaining temporal and cross-site variation (affine and residual, respectively). 

\subsubsection{CI Perturbation Selection}
\label{subsec:ci-perturbation-selection}
We select perturbations according to observed coverage, numerical differences, and disagreement structure using the rules in Table~\ref{tab:perturbation-selection}. Selection thresholds are fixed from characterisation data before the RMS experiments.
Table~\ref{tab:perturbations} summarises the candidate provider-switch, availability, and numerical perturbations. These vary the CI input while holding the workload, topology, constraints, and policy fixed to assess changes in placement decisions and operational outcomes.

\begin{table}[t]
\centering
\caption{Candidate CI perturbation types. Perturbation types are selected only when supported by the characterisation.}
\label{tab:perturbations}
\small
\begin{tabularx}{\columnwidth}{@{}l X@{}}
\toprule
Type & Definition \\
\midrule
Provider switch &
$\widetilde C_{z,t}=C^{(s')}_{z,t}$ for a matched observation from another provider $s'$. \\[1mm]

Additive &
$\widetilde C_{z,t}=\Pi(C_{z,t}+b_z)$, where $b_z$ is calibrated from
observed affine offsets. \\[1mm]

Multiplicative &
$\widetilde C_{z,t}=\Pi(a_zC_{z,t})$, where $a_z$ is calibrated from observed affine slopes. \\[1mm]

Empirical residual &
$\widetilde C_{z,t}=\Pi(C_{z,t}+r^{*}_{z,t})$, where $r^{*}$ is block-resampled from observed affine residuals while preserving temporal and, where possible, cross-zone dependence. \\[1mm]

Joint affine--residual &
$\widetilde C_{z,t}=
\Pi(\alpha_z+\beta_zC_{z,t}+r^{*}_{z,t})$, combining offset and scaling differences with the remaining residual disagreement. \\[1mm]

\bottomrule
\end{tabularx}
\end{table}

\subsection{CI Placement Sensitivity and Robustness}
\label{subsec:decision-level-ci-analysis}

A carbon-aware RMS uses CI information to decide \textit{where} and \textit{when} workloads should execute. The decision examined by the \escape framework therefore depends on the ``system under test''.
We apply the framework to CarbonEdge, as the relevant decision is the placement of workload across candidate sites. The Robustness/Sensitivity Analysis Engine component (Figure~\ref{fig:architecture-diagram}) compares placement and operational outcomes differences between multiple runs with fixed workload, topology, and energy profiles, while varying the CI.
We evaluate propagation at three levels. First, \emph{CI-ranking sensitivity} determines whether changing the CI source or applying a perturbation changes the ordering of candidate sites. Second, \emph{placement sensitivity} determines whether this change produces a different RMS decision. A \textit{placement flip} occurs when the complete placement produced for a matched workload batch differs from its reference placement; the \textit{flip rate} is the proportion of matched batches for which this occurs. Third, \emph{operational sensitivity} measures whether the changed placement affects carbon, energy, or latency.

\section{Experiment Design and Setup}
\label{sec:experimental-evaluation}

\subsection{Experimental Setup}
\label{subsec:evaluation-design}

We instantiate the \escape framework by reproducing CarbonEdge's RMS for the Central European region. We retain all the parameters from the original experiment, while only varying the CI-provider trace by applying its controlled perturbation. This design isolates the effects of changes in the CI input on placement decisions \textit{output} on carbon, energy, and latency \textit{outcomes}. 

First, the 2024 \emapshort and \wn traces are compared. Paired runs are then executed using each CI-provider trace, and the resulting placements and system outcomes are evaluated. Controlled CI perturbations to assess placement sensitivity are applied. The provider-switch and perturbation comparisons align repeated runs by date and batch checkpoint. Across runs, the topology, workload, latency matrix, server capacities, workload profile, placement policy, and all non-CI parameters are held constant. The CI trace is varied while application ordering and placement remain subject to simulator non-determinism. We distinguish four analytical levels of disagreement: (1) CI values, (2) candidate-site rankings, (3) the dominant placement site, and (4) complete workload placements.

\subsection{CI Data and Harmonisation}
\label{subsec:ci-data-harmonisation}

We use hourly CI observations for 2024 from \emapshort and \wn.
The experiment considers five placement sites: Graz (Austria, \texttt{AT}), Lyon (France, \texttt{FR}), Munich (Germany, \texttt{DE}), Milan (Italy, \texttt{IT-NO}), and Bern (Switzerland, \texttt{CH}). The geographical coverage of the analysis is therefore limited to these zones.

For each CI-provider-zone trace, timestamps are parsed as UTC and CI values are coerced to numerical form. Rows with invalid timestamps or CI values are removed, duplicate timestamps are reduced to one observation, and the remaining rows are ordered chronologically; conflicting duplicate values fail the validation pipe
-line. The \emapshort and \wn traces are then joined by site and timestamp. No missing CI values are imputed. Coverage and missingness are calculated from the source traces before this overlap-only join, whereas disagreement, candidate-site ranking differences, and affine and residual structure are calculated from matched observations.

\begin{figure}[t]
  \centering
  \resizebox{\columnwidth}{!}{%
    \input{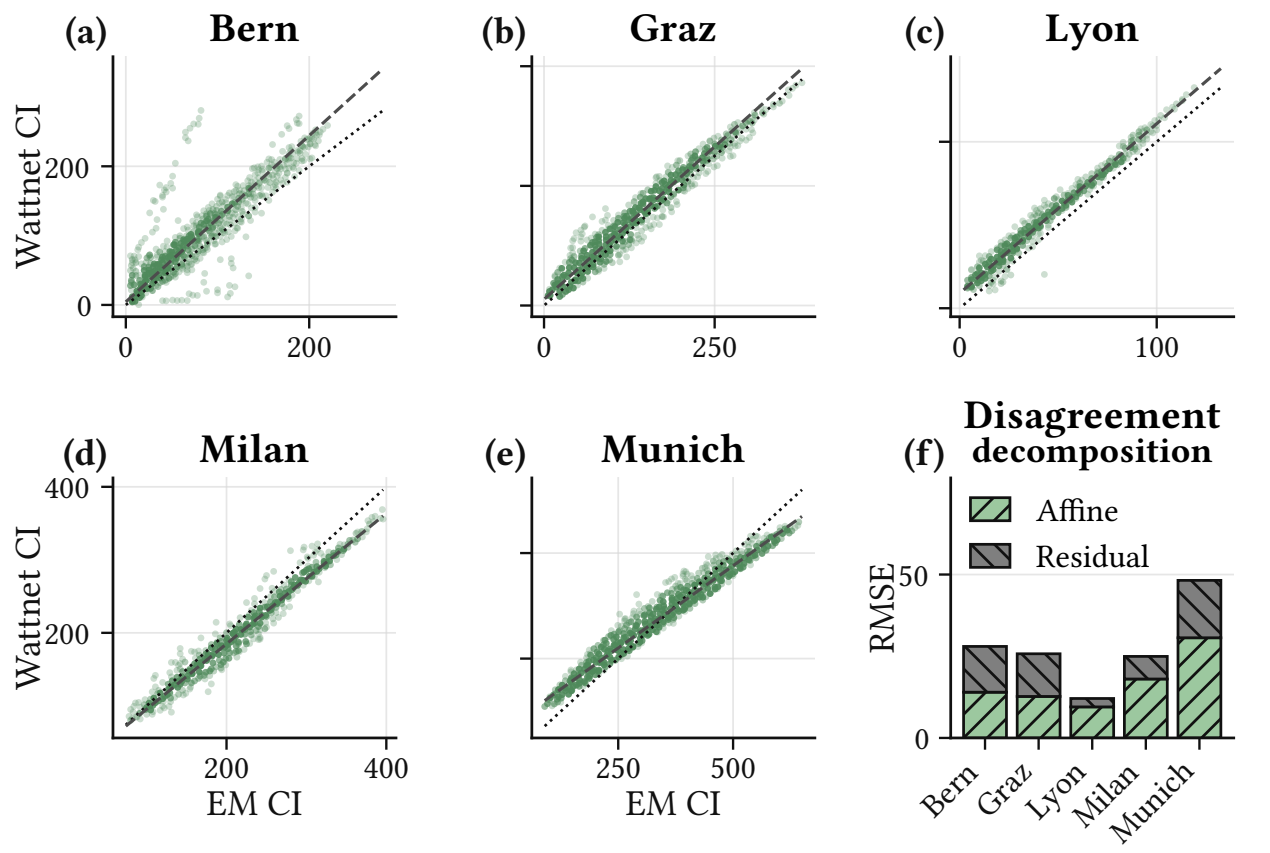}
  }
  \caption{CI disagreement-structure diagnostic.}
  \label{fig:ci-structure-diagnostics}
\end{figure}

Figure~\ref{fig:ci-structure-diagnostics} compares matched \emapshort and \wn observations across the five sites and distinguishes affine from residual disagreement. The observed provider differences also parameterise the empirical CI perturbations.

\subsection{Simulator, Topology, and Workload}
\label{subsec:simulator-topology-workload}

We reproduce the CarbonEdge online simulator and its placement policies using Google OR-Tools. Placement feasibility requires the total latency, including directed network Round-Trip Time (RTT), input-transfer time, and queue-adjusted inference latency; to remain within the application Service-Level Objectives (SLOs) while satisfying GPU and CPU utilisation, memory and bandwidth contraints.
Following the Central European scenario evaluated in the original CarbonEdge paper~\cite{wuCarbonEdgeLeveragingMesoscale2025b}, the topology comprises five sites (Bern, Graz, Lyon, Milan, and Munich), three homogeneous NVIDIA A2 servers per site, and therefore 15 servers in total.
The workload comprises six ResNet50 application instances originating from each site, yielding 30 applications per daily run. The applications are placed sequentially across 11 online batches at a request rate of 10 requests/s. For ResNet50 on an NVIDIA A2 GPU, the simulator uses a mean inference latency of $2.952\,\mathrm{ms}$, GPU-memory demand of $118\,\mathrm{MB}$, and device power of $29.642\,\mathrm{W}$. The simulator converts each site’s hourly CI observations into a 24-hour vector and uses its daily mean in the placement objective.

\subsection{Policies and Reference Baselines}
\label{subsec:policies-baselines}

We evaluate four policy families: \texttt{Carbon-aware}, \texttt{Intensity-aware}, \texttt{Energy-aware}, and \texttt{Latency-aware}. Latency-aware placement uses the greedy implementation, using one implementation per family. \texttt{Intensity-aware} and \texttt{Carbon\--aware} placements use CI directly, whereas \texttt{Energy\--aware}, and \texttt{La\-tency\--aware} placements provide reference policies whose objectives do not explicitly depend on CI. For each policy, multiple experiments hold the same workload, topology, resource capacities, latency constraints, and all other parameters constant, varying only the CI input. Provider-specific estimates are averaged with equal weights across \emapshort and \wn~(Figure \ref{fig:perturbation-sensitivity}).

Provider-switch results are reported for \texttt{Intensity-aware} placement. In the perturbation analysis, flip-rates and carbon regret are reported separately for all four policies. Energy and latency percentage changes are first calculated relative to each provider-specific control and then averaged equally across providers. The bottom figure row assigns weight \(1/2\) to Latency-aware and \(1/6\) to each remaining policy. \texttt{Carbon-aware} placement provides the reference for carbon-regret calculations.

\subsection{CI-Provider Switches and Perturbation Sweeps}
\label{subsec:perturbation-protocol}

We do not modify CI values when switching between \emapshort and \wn CI-providers traces; we supply each trace to otherwise identical CarbonEdge runs. Non-negative clipping is applied after every constructed numerical perturbation.
We analyse positive and negative CI perturbations separately, as fluctuations may affect placement decisions differently. The empirical noise sweep samples, with replacement, from the observed \wn--\emapshort differences within each site. It does not preserve temporal autocorrelation, including discrepancies that persist across consecutive hours.
We run each unperturbed simulation five times and each perturbation setting five times, aggregating metrics within each run. We report differences between perturbed and unperturbed mean operational outcomes, and baseline-corrected placement flip-rates relative to disagreement between unperturbed repeats.

\subsection{Evaluation Metrics}
\label{subsec:evaluation-metrics}

We evaluate CI sensitivity through the following metric categories:
\textit{(1) Signal characterisation} measures coverage through spatial availability, temporal completeness, and pairwise overlap. Matched CI values are compared using bias, Mean Absolute Error (MAE), Root Mean Squared Error (RMSE), and Pearson correlation; affine fits and residual statistics characterise disagreement structure.
\textit{(2) CI-ranking and placement sensitivity} is assessed using top-ranked-site disagreement, top-three overlap, effective margin-split, and Spearman correlation on daily mean CI rankings. Top-ranked CI candidate disagreement measures how often the providers identify different lowest-CI sites. Dominant-site disagreement measures how often the site receiving the most application assignments differs. The raw complete placement flip-rate measures how often at least one allocation differs between the compared placements. We correct placement flip rates for variability between repeated runs. For provider switching, we subtract the average within-provider disagreement from cross-provider disagreement; for perturbations, we subtract each provider’s own baseline before averaging equally across providers. For each matched decision, we calculate the margin between the lowest- and second-lowest CI sites for each provider, and take the lowest gap. We compare the distribution of effective margins between decisions with changed and unchanged complete placements after switching providers.
\textit{(3) Perturbation sensitivity} the primary metrics is placement flip-rate, which is assessed by comparing each perturbed run with its unmodified provider-specific reference. Placement flip-rates are reported by perturbation type, magnitude degree, and policy.
\textit{(4) System-level effects} are measured through date- and step-aligned differences in the carbon proxy, latency, and energy (where positive values indicate increases relative to the reference); and the carbon regret metric, which measures excess carbon relative to the Carbon-aware (CarbonEdge's use case) policy under the same CI input. We report the mean difference across matched steps; positive values indicate a higher carbon proxy than the reference.

\section{Results and Analysis}
\label{sec:results-analysis}

\subsection{CI Source Coverage, Numerical Alignment, and Disagreement-Structure}
\label{subsec:ci-source-comparison}
 
For the characterisation step we consider coverage, numerical alignment, and provider disa\-greement-structure.
The fixed 2024 analysis window contains \(43{,}920\) expected site--hour observations across the five Central European sites. EM contains all \(43{,}920\) expected observations, whereas \wn contains \(43{,}202\), corresponding to \(98.37\%\) temporal completeness. Both pro\-viders cover all five sites, but \wn has lower temporal completeness because 718 hourly observations are missing for Bern; the other four sites have complete coverage. The harmonised comparison therefore contains \(43{,}202\) matched site--hour pairs, with no duplicate site--timestamp keys and no missing CI values after the inner join, as summarised in Table~\ref{tab:ci-coverage-results}.

\begin{table}[t]
  \centering
  \caption{Temporal completeness within the five-site experimental scope. Both \wn (\wnshort) and ElectricityMaps (\emapshort) contain observations for all five sites.}
  \label{tab:ci-coverage-results}
  \small
  \begin{tabular}{lrrr}
    \toprule
    Source/site & Present & Expected & Completeness \\
    \midrule
    EM, all sites & 43,920 & 43,920 & 100.00\% \\
    WN, all sites & 43,202 & 43,920 & 98.37\% \\
    \midrule
    WN, Bern   & 8,066 & 8,784 & 91.83\% \\
    WN, Graz   & 8,784 & 8,784 & 100.00\% \\
    WN, Lyon   & 8,784 & 8,784 & 100.00\% \\
    WN, Milan  & 8,784 & 8,784 & 100.00\% \\
    WN, Munich & 8,784 & 8,784 & 100.00\% \\
    \bottomrule
  \end{tabular}
\end{table}

Table~\ref{tab:ci-source-alignment} summarises CI-source alignment by site. Across the ma\-tched sample, the mean \wn--\emapshort difference is \(11.020\)\carbonunit, the aggregate MAE and RMSE are \(14.785\)\carbonunit and \(20.433\)\carbonunit, respectively; and the pooled Pearson correlation is \(0.990\). This indicates strong temporal co-movement, meaning that both providers often capture similar fluctuations. However, the observed disagreement and zone-specific biases show that their absolute CI values are not numerically interchangeable.
Disagreement is spatially heterogeneous. For example, Lyon has the lowest MAE, at \(6.864\)\,\carbonunit, whereas Munich has the highest, at \(26.547\)\,\carbonunit. \wn is higher on average in Bern, Graz, Lyon, and Munich, but lower in Milan.

\begin{table*}[t]
\centering
\caption{CI-source numerical alignment and disagreement structure by site.
\(n\) denotes matched observations; bias is \wn minus \emapshort.
The affine model is \(\mathrm{WN}=\alpha+\beta\,\mathrm{EM}\), where
\(\alpha\) is the offset and \(\beta\) the scaling factor.
\(r\) denotes Pearson correlation.
\(\mathrm{RMSE}_{\mathrm{affine}}\) measures the fitted affine component
of disagreement, and \(\mathrm{RMSE}_{\mathrm{residual}}\) measures
the remaining differences. CI values, offsets, and disagreement measures
are expressed in \carbonunit.}
\label{tab:ci-source-alignment}
\scriptsize
\setlength{\tabcolsep}{3.0pt}
\begin{tabular}{@{}lrrrrrrrrrrr@{}}
\toprule
& & \multicolumn{2}{c}{Mean CI}
& \multicolumn{3}{c}{Raw disagreement}
& \multicolumn{3}{c}{Affine fit and correlation}
& \multicolumn{2}{c}{Disagreement components} \\
\cmidrule(lr){3-4}
\cmidrule(lr){5-7}
\cmidrule(lr){8-10}
\cmidrule(l){11-12}
Site & \(n\)
& EM & WN
& Bias & MAE & RMSE
& \(\beta\) & \(\alpha\) & \(r\)
& \(\mathrm{RMSE}_{\mathrm{affine}}\)
& \(\mathrm{RMSE}_{\mathrm{residual}}\) \\
\midrule
Munich & 8,784 & 279.28 & 299.98 & 20.70 & 26.55 & 32.38
& 0.832 & 67.68 & 0.984 & 27.88 & 16.46 \\
Graz & 8,784 & 79.99 & 105.26 & 25.26 & 25.28 & 27.23
& 1.030 & 22.85 & 0.988 & 25.33 & 9.99 \\
Bern & 8,066 & 17.00 & 22.79 & 5.78 & 7.46 & 12.00
& 1.413 & -1.23 & 0.976 & 10.14 & 6.41 \\
Milan & 8,784 & 172.93 & 169.01 & -3.93 & 7.17 & 9.04
& 0.885 & 15.97 & 0.992 & 7.05 & 5.67 \\
Lyon & 8,784 & 15.42 & 22.28 & 6.86 & 6.86 & 7.06
& 0.967 & 7.37 & 0.990 & 6.87 & 1.63 \\
\bottomrule
\end{tabular}
\end{table*}

The site-specific affine model explains between \(95.2\%\) and \(98.4\%\) of \wn variation. This means that much of the provider disagreement is systematic, following a persistent pattern offset and scaling that is site-specific. The two sources often preserve temporal structure, but differ in level and scale. However, the RMSE of the differences remain after adjusting for site-specific offsets and scaling remains non-negligible, ranging from \(1.634\)\carbonunit in Lyon to \(16.456\)\,\carbonunit in Munich. The residual correlations between consecutive hours range from \(0.738\) to \(0.948\) indicates that disagreement persists over time even after adjusting for offsets and scaling.

\subsection{Provider-Switch Impact}
\label{subsec:provider-switch-results}

This experiment compares matched \emapshort and \wn CI-source runs under the same topology, workload, latency matrix, server capacities, workload profile, and policy (\texttt{Intensity-aware}).

\begin{figure}[t]
  \centering
  \resizebox{\columnwidth}{!}{%
    \input{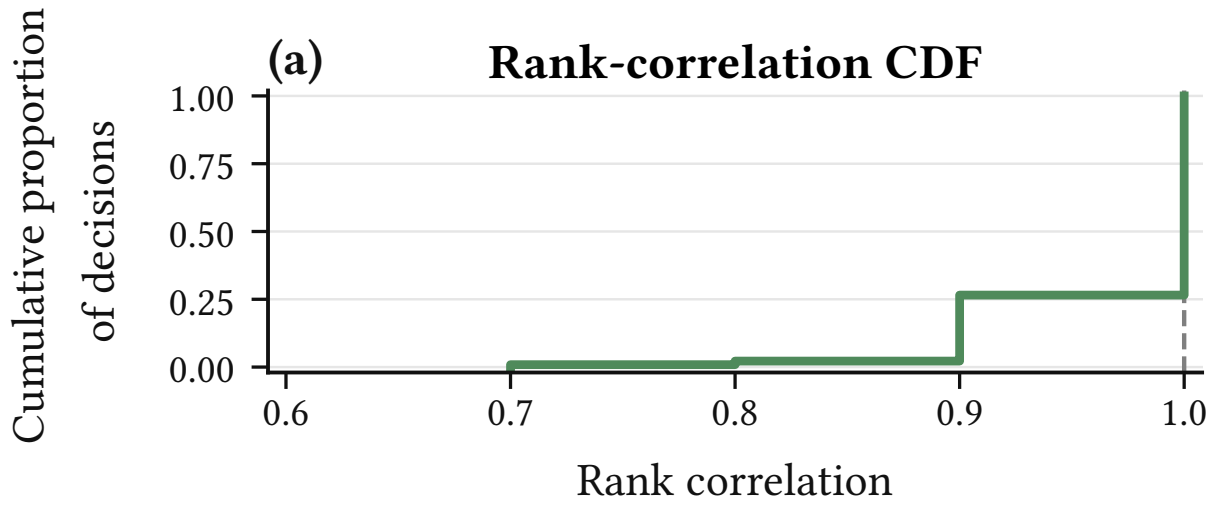}%
  }
  \caption{CI-ranking agreement between EM and WN. The rank-correlation Cumulative Distribution Function (CDF) shows that most matched decisions preserve identical or near-identical CI candidate-site rankings.}
  \label{fig:source-switch-decision-impact}
\end{figure}

Table~\ref{tab:source-switch-impact} summarises the \textit{decision-level} effects of provider switching, Figure~\ref{fig:source-switch-decision-impact} shows the distribution of CI-ranking correlations, and Table~\ref{tab:source-switch-outcomes} summarises the operational outcomes. The paired decision table contains \(2{,}486\) aligned decision steps per run pair, and \(62{,}150\) comparisons across 25 cross-provider run pairs. Replacing \emapshort with \wn changes the top-ranked CI site in \(14.60\%\) of matched placement decisions. Despite this, the mean Spearman rank correlation remains high at \(0.970\), and the top-three overlap of \(98.23\%\). Dominant-site and complete placement disagreement reach \(43.50\%\) and \(77.14\%\), respectively. The baseline-corrected placement flip-rate is \(-0.93\) percentage points, relative to a symmetric within-provider baseline of \(78.07\%\). Median effective CI margins are \(4.46\) for changed placements and \(5.46\) for unchanged placements. Changed placements have an approximately \(18\%\) lower median effective CI margin compared to unchanged placements, indicating that placement changes are associated with more closely ranked CI alternatives.
Across 226 paired days, switching providers increases the mean operational carbon proxy by \(15.07\%\), while mean latency and energy use decrease by \(0.48\%\) and \(1.02\%\), respectively. The carbon-proxy difference reflects both placement changes and provider-specific CI accounting. Frequent placement changes therefore coexist with small average changes in energy and latency; a placement flip alone does not necessarily establish an operational penalty or worse outcome.

\begin{table}[t]
\centering
\caption{CI-ranking and placement disagreement between EM and WattNet under the Intensity-aware policy. Comparisons are aligned by date and batch checkpoint. The symmetric baseline averages within-provider disagreement across the ten distinct run pairs for each provider. The baseline-corrected flip-rate is the cross-provider flip-rate minus this baseline. pp denotes percentage points. Effective CI margins are in $\mathrm{gCO_{2}eq/kWh}$ and reported as median [IQR].}
\label{tab:source-switch-impact}
\small
\setlength{\tabcolsep}{3pt}
\begin{tabularx}{\columnwidth}{@{}Xr@{}}
\toprule
Metric & Value \\
\midrule
Complete paired days & 226 \\
Matched decision steps per run pair & 2\,486 \\
Top-ranked CI candidate disagreement & 14.60\% \\
Dominant-site disagreement & 43.50\% \\
Complete placement flip-rate & 77.14\% \\
Baseline-corrected placement flip-rate & $-0.93$ pp \\
Mean rank correlation & 0.970 \\
Top-three candidate overlap & 98.23\% \\
Effective CI margin, unchanged placement
    & 5.46 [2.00, 12.00] \\
Effective CI margin, changed placement
    & 4.46 [1.53, 12.22] \\
\bottomrule
\end{tabularx}
\end{table}

\begin{table}[t]
    \centering
    \small
    \setlength{\tabcolsep}{3pt}
    \caption{Mean daily operational outcomes across 226 aligned days
    and five runs per provider under the Intensity-aware policy.
    Changes are relative to EM when switching to \wn,
    calculated from the provider means.}
    \label{tab:source-switch-outcomes}
    \begin{tabularx}{\columnwidth}{@{}Xrrr@{}}
        \toprule
        Metric & EM & \wn & Change (\%) \\
        \midrule
        Carbon proxy ($\mathrm{g\cdot CO_2eq/day}$)
            & 129.40 & 148.90 & $+15.07$ \\
        Energy ($\mathrm{kWh/day}$)
            & 1.2957 & 1.2825 & $-1.02$ \\
        Latency ($\mathrm{ms}$)
            & 12.7540 & 12.6924 & $-0.48$ \\
        \bottomrule
    \end{tabularx}
\end{table}

\subsection{Perturbation Sensitivity}
\label{subsec:perturbation}

\begin{figure*}[t]
  \centering
  \resizebox{\textwidth}{!}{%
    \input{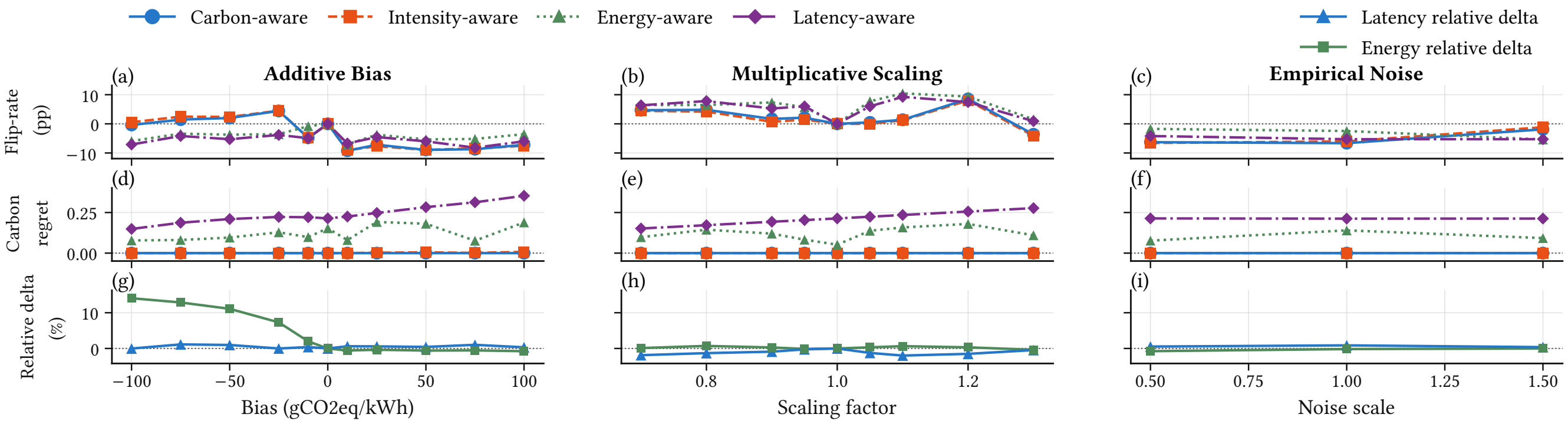}%
  }
  \caption{Perturbation sensitivity averaged equally across EM and
    WattNet. The top row reports baseline-corrected flip-rate in percentage points. The middle row reports carbon regret relative to Carbon-aware placement under the same perturbed CI input, in gCO$_2$ per decision checkpoint. The bottom row reports weighted mean energy and latency percentage changes relative to provider-specific controls, using the policy weights described in Section~\ref{subsec:policies-baselines}.}
  \label{fig:perturbation-sensitivity}
\end{figure*}

Figure~\ref{fig:perturbation-sensitivity} summarises placement sensitivity to CI perturbations. In the results averaged equally across \emapshort and \wn, flip-rate varies with perturbation magnitude and direction. Sensitivity varies across perturbation directions and individual policies; for example, an additive bias of \(+10\)\,\carbonunit produces a baseline-corrected flip-rate of \(-6.73\) percentage points for the Latency-aware policy. These results describe observed differences at the tested magnitudes; they do not identify the minimum perturbation required to alter placement.

Carbon regret is consistently higher for Latency-aware placement than for the other evaluated policies, and generally increases with positive additive bias and larger multiplicative factors. However, policy-specific flip-rates show no consistent (monotonic) response to perturbations. Operational effects also depend on the perturbation type; for example, an additive bias of \(-100\)\carbonunit increases weighted mean energy use by \(14.06\%\), while weighted mean latency changes by approximately \(-0.02\%\). Multiplicative perturbations produce smaller, but non-monotonic changes in both metrics. Under empirical noise, carbon regret remains approximately constant and weighted mean energy and latency changes remain close to zero across the tested scales, despite changes in placement. These averages may conceal opposing provider-specific responses and differences between policies. These results indicate that placement sensitivity need not translate into comparable changes in average operational outcomes.

\subsection{Discussions and Limitations}
\label{subsec:discussions-limitations}

Strong temporal correlation between \emapshort and \wn does not imply that their CI estimates are interchangeable. These estimates differ in level, scale, and residual variation, with heterogenous differences across sites. During RMS design phase, this means that agreement should be checked across target candidate locations, as neither CI source constitutes the ``ground-truth''. Their disagreement provides a proxy for uncertainty, which impacts placement decisions and operational outcomes.

Across \emapshort and \wn runs, dominant sites differ in \(43.50\%\) of aligned decisions despite a mean CI-ranking correlation of \(0.970\). The baseline-corrected placement flip-rate is \(-0.93\) percentage points, indicating that the raw cross-provider flip-rate is similar to the symmetric within-provider baseline. Nevertheless, mean energy use and latency decrease by only 1.02\% and 0.48\%, respectively; illustrating why ranking changes, placement flips, and operational outcomes must be distinguished. The operational carbon proxy increases by 15.07\%, but each run uses its own provider's CI, so this difference combines placement effects with changes in emissions accounting. Isolating the carbon consequences of placement therefore requires evaluating both allocations under a common CI reference.

Controlled perturbations provide a practical way to examine how strongly an RMS depends on its CI input, and how its uncertainty impacts placement decisions. RMS evaluation should therefore test different types and directions of CI variation, and compare placement changes along with the associated energy, carbon, latency, and other operational outcomes differences. Such tests can also support the evaluation of policies intended to accommodate uncertainty.

\section{Conclusion and Future Work}
\label{sec:conclusion}

Carbon-aware RMSs rely on external CI signals whose estimation and forecasting \textit{uncertainty} can make their decisions \textit{fragile}. We instantiated \escape using \emapshort and \wn as comparative CI-source baselines and applied it to a reproduced CarbonEdge mesoscale placement RMS under fixed workload, topology, latency, resource, and energy conditions. Across \(2,486\) aligned decision steps per run pair over 226 days, switching from \emapshort to \wn changed the top-ranked CI site in 14.60\% of decisions, with dominant-site disagreement in \(43.50\%\) of cross-provider comparisons, despite a mean rank correlation of \(0.970\). Mean energy use and latency decreased by 1.02\% and 0.48\%, respectively, while the operational carbon proxy increased by 15.07\%, reflecting differences in both placement and provider-specific CI accounting.

The baseline-corrected placement flip-rate is \(-0.93\) percentage points, indicating a raw cross-provider flip-rate similar to the symmetric within-provider baseline. Placement differences cannot therefore be attributed to CI-source choice alone. CI-source choice affects reported operational outcomes, even when aggregate CI rankings remain highly aligned. Frequent placement changes can coexist with small average differences in energy use and latency, showing that ranking stability, placement stability, and operational outcomes must be evaluated separately. RMS solutions should therefore account for CI uncertainty during both design and evaluation.

\bibliographystyle{ACM-Reference-Format}
\bibliography{references}

\appendix
\end{document}